\documentclass[
superscriptaddress,
preprint,
showpacs,
amsmath,amssymb,aps,
pra,
]{revtex4-1}

\usepackage{graphicx}
\usepackage{dcolumn}
\usepackage{bm}
\usepackage{hyperref}

\begin{document}

\preprint{}

\title{Interference between two independent multi-mode thermal fields}

\author{Jie Su}
\author{Jiaming Li}
\author{Liang Cui}

\author{Xiaoying Li}%
\email{xiaoyingli@tju.edu.cn} 
\affiliation{School of Precision Instrument and Opto-electronics Engineering, Tianjin University, Key Laboratory of Opto-electronic Information Technology of Ministry of Education, Tianjin 300072, China}

\author{Z. Y. Ou}%
\affiliation{School of Precision Instrument and Opto-electronics Engineering, Tianjin University, Key Laboratory of Opto-electronic Information Technology of Ministry of Education, Tianjin 300072, China}
\affiliation{Department of Physics, Indiana University-Purdue University Indianapolis, 402 N. Blackford St., Indianapolis, Indiana 46202, USA}

\date{\today}

\begin{abstract}
We study the property of the field which is a mixing of two multi-mode thermal fields. We accomplish a general theoretical analysis and show that the mode of the mixed field, characterized by its intensity correlation function $g^{(2)}$, is determined by the two-photon interference between the two independent multi-mode thermal fields. Our analysis reveals that the mode structures of the two thermal fields play an important role in the interference. Comparing with $g^{(2)}$ for one of the individual field with less average mode number, $g^{(2)}$ of the mixed field always decreases due to the change of mode distribution, but the amount of drop depends on the relative overlap between the mode structures of the two thermal fields and their relative strength. Moreover, we verify the theoretical analysis by performing the experiments when the modes of two multi-mode thermal fields are identical, orthogonal and partially overlapped, respectively. The experimental results agree with theoretical predictions. Our investigation is useful for analyzing the signals carried by the intensity correlation of multi-mode thermal fields.
\end{abstract}

\maketitle


\section{\label{sec:level1}Introduction}

In nature, most light sources are of thermal nature because of the randomness of individual emitters \cite{loudon}. Therefore, what we see often is mostly the addition of many thermal sources. When two independent thermal fields are mixed, one usually thinks that there is no interference between them and the result is simply the addition of the two. This is true at intensity level and since background noise is often of the nature of thermal source, we can simply subtract them out. However, this is not true at higher order measurement such as intensity correlation.

It was shown as early as in 1967 \cite{pflee} that when two independent fields are mixed, although intensity shows no interference, intensity correlations do give rise to interference patterns, which is known as fourth-order interference or two-photon interference. This problem of mixing two independent thermal sources becomes prominent when the optical signals need to be extracted from the intensity correlation of thermal fields. For example, in the technique of ghost imaging \cite{ghost} where higher order intensity correlations of thermal sources are used,  the access correlations beyond accidental serve as the signals to extract the image. So, if both the background noise and the signal fields are in thermal states, the effect of mixing will show up as interference in access correlation and the background cannot be easily taken out. Moreover, the access correlations from a thermal source are from photon bunching effect first observed by Hanbury-Brown and Twiss (HBT) \cite{HBT}. Thus the ghost imaging technique would prefer to have the photon bunching effect as large as possible for good signal-to-noise ratio (SNR) in imaging. But the interference between the signal and background noise may affect the SNR extracted from the fourth-order interference effect.

Two-photon interference between independent thermal sources was first studied by Mandel in a seminal paper \cite{man83} on interference in intensity correlation. It has been shown that the visibility of two-photon interference between two independent thermal sources has a maximum value of 1/3. But most discussion on mixing of thermal sources are based on single-mode models under the ideal conditions which give rise to the maximum photon bunching effect \cite{Ou97}. As is known, however, the multi-mode nature of the thermal sources will reduce the photon bunching effect. It will certainly affect the interference effect between two thermal sources and the photon bunching effect in the mixed field.

The mode properties of optical fields, however, are sometimes complicated and can not be characterized easily. This is because the two fields participating in interference such as the signal and the background fields may have different mode structures and the interference between fields with different mode structures will lead to the change of the mode distribution. In this paper, we will study how mode structures of the thermal fields affect the two-photon interference between them. It is found that the intensity correlation function $g^{(2)}$ will change after the mixing, depending on the relative overlap between the mode structures of the two fields and their relative strength. Moreover, we perform experiments to verify the theoretical prediction.

The rest of paper is organized as follows. In section II, we theoretically study the influence of the mode structure of the two thermal fields on the two-photon interference effect shown in mixed field. By developing a general theoretical model, the formula for describing the upper bound and lower limit of $g^{(2)}$ for the mixed field are deduced. In section III, we describe the experimental verification, in which the interference effect of two multi-mode thermal fields with identical mode structure and different mode structures are measured and analyzed. Conclusions and discussion are presented in the last section.



\section{The general theoretical analysis}
The mode number of a multi-mode thermal field, reflected by its photon statistics, can be characterized by normalized intensity correlation function. Beginning with a brief review of the photon statistics for a thermal field, we will study the mode property of the mixed field and demonstrate that the two-photon interference between the two independent multi-mode thermal fields play an important role. Moreover, we will analyze the factors influencing the interference.

\subsection{\label{sec:level1} Photon statistics for a thermal field in multiple modes}
A thermal light field is a random process with complex Gaussian probability distribution \cite{loudon}. Its description, however, depends on the mode structure we use. For the stationary field of continuous wave, a common approach of distinguishing the modes is by frequency \cite{mandel-wolf}. Since the frequency and time are conjugate variables, a set of overlapping but orthogonal broadband wave-packet modes, named as ``temporal modes", can also forms a complete mode basis. In this paper, we adopt the temporal modes (TMs) \cite{brecht15} to describe pulsed thermal fields. For thermal lights confined by the waveguide with single-transverse mode, such as optical fiber, the TMs form a complete basis for representing an arbitrary optical field.

\begin{figure}[!htp]
	\includegraphics[width=8.5cm]{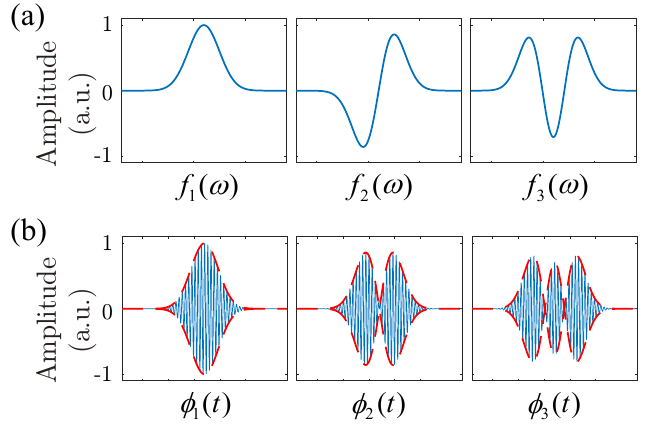}
	\caption{\label{fig:1} The first three Hermite-Gaussian modes of a temporal mode basis in (a) frequency domain and (b) time domain, respectively.}
\end{figure}

For a thermal optical field propagating in single-mode optical fibers, one dimensional approximation applies. In this case, the electric field amplitude of a thermal field  at time $t$ can be written as
\begin{equation}\label{eq:Amplitude}
E(t) = \vec{e} \sum_i A_i\phi_i(t),
\end{equation}
where $\vec{e}$ is a unit polarization vector, $A_i$ is a random variable with Gaussian statistics for mode $i$, $\phi_i(t)=\frac{1}{2\pi}\int f_i(\omega)e^{-\text{i} \omega t} d\omega$ is the temporal mode profile for mode $i$ with continuous spectrum $f_i(\omega)$. To demonstrate TMs form a family of Hermite-Gaussian functions of frequency, we exemplarily plot the first three members of a TM basis in Fig.~\ref{fig:1}. Although the TMs can be fully overlapped in polarization, space, frequency, and time, TMs are orthogonal with respect to a frequency (time) integral:
\begin{equation}\label{}
\int\phi^*_i(t)\phi_j(t) dt = \frac{1}{2\pi} \int f^*_i(\omega) f_j(\omega) d\omega=\delta_{i,j}.
\end{equation}
In Eq.~(\ref{eq:Amplitude}), we have $\langle A_i \rangle = 0$ and $\langle A_i^* A_j\rangle = 0\;(i\!\neq\! j)$ for $i,j = 1,2,\cdots$, because of the phase randomness of thermal radiation, while the intensity distributed in mode $\phi_i(t)$ is described by $\langle |A_i|^2 \rangle = \alpha_i^2$. 

\begin{figure}[!htp]
	\includegraphics[width=9.5cm]{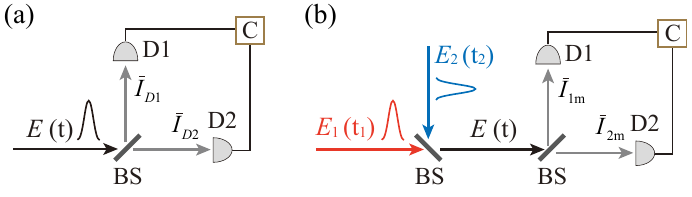}
	\caption{\label{fig:2} The schematic of Hanbury Brown-Twiss (HBT) interferometer for intensity correlation measurements of (a) a thermal field and (b) a mixing of two individual thermal fields, respectively. BS, 50/50 beam splitter; D, detector; C, correlator.}
\end{figure}

The photon statistic of the thermal field is characterized by the HBT interferometer consisting of a 50/50 beam splitter (BS) and two detectors (D1 and D2), as shown in Fig.~\ref{fig:2}(a). When the response time of each detector is much longer than the coherence time of thermal field, the average intensity of thermal field measured by each detector (D1 and D2) \cite{goodman} is 
\begin{equation}\label{eq:intensity}
\begin{aligned}
\bar I_{D1(2)}&=\left\langle I_{D1(2)} \right\rangle\\
&=\left\langle \int I_{D1(2)}(t) dt \right\rangle\\
&= \left\langle \int E_{D1(2)}^*(t) E^{}_{D1(2)}(t)dt \right\rangle\\
&= \frac{1}{2} \left\langle \int E^*(t)E(t)dt\right\rangle,
\end{aligned}
\end{equation}
where $E^{}_{D1(2)}(t) = E(t)/\sqrt{2}$. Substituting Eq.~(\ref{eq:Amplitude}) into Eq.~(\ref{eq:intensity}), we arrive at
\begin{equation}\label{eq:aveintensity}
\begin{aligned}
\bar I_{D1(2)}=\frac{1}{2} \Big[\sum_{i,j}  \left\langle  A_i^* A_j \right\rangle  \int \phi^*_i(t) \phi_j(t) dt \Big] = \frac{1}{2} \sum_{i} \alpha_i^2.
\end{aligned}
\end{equation}
The second-order intensity correlation function measured by the two detectors is
\begin{equation}\label{eq:2ndICF}
\begin{aligned}
\left\langle I_{D1}I_{D2} \right\rangle 
&=\left\langle \int  I_{D1}(t)I_{D2}(t') dt dt{'}  \right\rangle\\
&=\left\langle \int E_{D1}^*(t)E_{D2}^*(t') E^{}_{D2}(t') E^{}_{D1}(t) dt dt{'}  \right\rangle\\
&=\frac{1}{4}\Big[\sum _i \alpha_i^4 + \big(\sum \alpha_i^2\big)^2\Big].
\end{aligned}
\end{equation}
After normalizing the intensity correlation with the average intensities measured by D1 and D2, we then arrive at
\begin{equation}\label{eq:singleg2}
g^{(2)} = \frac{\left\langle I_{D1}I_{D2} \right\rangle}{\bar I_{D1} \bar I_{D2}}
=1 + \frac{\sum_{i}\alpha_i^4}{\big(\sum_i \alpha_i^2\big)^2} =1+ {1\over M},
\end{equation}
where $M\equiv (\sum_i \alpha_i)^2/\sum_{i}\alpha_i^4$ is the average mode number of field $E(t)$. For a single-mode field, we have $M=1$. In this case, $g^{(2)}$ has the maximum value of 2, which corresponds to the maximum photon bunching effect of thermal field. Moreover, if we assume the intensity of the field is equally distributed in each mode, i.e., $\alpha_i^2=\alpha^2=I_0$ 
for $i=1,\cdots, M$, where $M$ is the number of modes, the average intensity in Eq.~(\ref{eq:aveintensity}) can be approximated as $\bar I_{D1(2)} = \frac{1}{2}MI_0$.

\subsection{\label{sec:level2} Photon statistics for a mixing of two multi-mode thermal fields}
Now we study the property of the thermal field, which is a mixing of two independent multi-mode thermal fields. As shown in Fig.~\ref{fig:2}(b), the mixed field $E(t)$ obtained by combining two thermal fields $E_1$ and $E_2$ with a 50/50 BS is written as
\begin{equation}\label{}
E(t) = \big[E_1(t_1) + E_2(t_2)\big] \big/ \sqrt{2},
\end{equation}
with
\begin{equation}\label{}
E_1(t_1) = \vec{e}_1  \sum_i A_i\phi_i(t_1),
\end{equation}
\begin{equation}\label{}
E_2(t_2) = \vec{e}_2  \sum_k B_k\varphi_k(t_2)
= \vec{e}_2  \sum_k B_k\varphi_k(t_1+\tau),
\end{equation}
where $\varphi_k(t)=\frac{1}{2\pi}\int g_k(\omega)e^{-\text{i} \omega t} d\omega$ is the temporal mode profile for mode $k$ of thermal field $E_2$, $\vec{e}_{1(2)}$ is the unit polarization vector, and $\tau=t_2-t_1$ denotes the delay between the two fields when they combined at BS. Because $A_i$ and $B_k$ are the complex Gaussian random variables, the following relations hold:
\begin{equation}\label{}
\begin{aligned}
\langle A_i \rangle = \langle B_k &\rangle = 0, \langle A_i^* B_k \rangle = \langle B_i^* A_k \rangle = 0,\\
\langle A^*_i A_j\rangle& = \delta_{i,j}\alpha_i^2, \langle B^*_k B_l \rangle = \delta_{k,l}\beta_k^2.\\
\end{aligned}
\end{equation}
The overlap for the TMs structure of $E_1$ and $E_2$ is described by the integral:
\begin{equation}\label{}
\int\phi^*_i(t)\varphi_k(t+\tau) dt = K_{i,k} \leqslant 1.
\end{equation}
When the mode bases of $E_1$ and $E_2$ are perfectly overlapped, we have $K_{i,k}=\delta_{i,k}$ ($\phi_i(t)=\varphi_i(t)$).

The instantaneous intensity of the mixed field is given by
\begin{eqnarray}\label{}
I_m(t)&= & E^*(t)E(t)\cr
&=&\Big[|E_1(t)|^2 \!+\! |E_2(t+\tau)|^2+ \cos\theta E_1^*(t)E_2(t+\tau) \cr &&\hskip 0.5in +\cos\theta E_2^*(t+\tau)E_1(t)\Big]\Big / 2,
\end{eqnarray}
where $\cos\theta = \vec{e}_1\cdot\vec{e}_2$ describes the overlap of the polarization modes between $E_1$ and $E_2$. When the photon statistics of the mixed field $E(t)$ is characterized by using HBT interferometer, as shown in Fig.~\ref{fig:2}(b), the average intensity measured by D1 or D2 is expressed as
\begin{equation}\label{}
\begin{aligned}
\bar I_{1m(2m)} &= \left\langle I_{1m(2m)} \right\rangle\\
& = {1\over 2}\left\langle \int I_m(t) dt \right\rangle\\
& = {1\over 4}\Big\{\sum_{i} \left\langle  |A_i|^2\right\rangle + \sum_{k} \left\langle  |B_k|^2\right\rangle\Big\} \\
& = {1\over 4}\Big(\sum_{i} \alpha_i^2 + \sum_{k} \beta_k^2\Big).
\end{aligned}
\end{equation}
Similar to Eq.~(\ref{eq:2ndICF}), the second-order intensity correlation function can be deduced as
\begin{equation}\label{}
\begin{aligned}
\left\langle I_{1m}I_{2m} \right\rangle &=
{1\over 4} \left\langle \int I_{m}(t)I_{m}(t') dt dt' \right\rangle\\
&=\frac{1}{16}\left\langle \sum_{i,k}(|A_i|^2+|B_i|^2)(|A_k|^2+|B_k|^2)\right\rangle\\
&\quad+\frac{1}{16}\Big[(\cos\theta)^2\left\langle \sum_{i,j,k,l}A^*_iB^*_kA_jB_lK^*_{ik}K_{jl}\right\rangle+c.c.\Big]\\
&\!= {1\over 16}\Big(\sum_{i} \alpha_i^2 + \sum_{k} \beta_k^2\Big)^2\!\\
&\quad+\!\frac{1}{16}\Big[\sum_{i}\alpha_i^4\!+\!\sum_{k}\beta_k^4\!+\!2(\cos\theta)^2\sum_{i,k}\alpha_i^2\beta_k^2|K_{ik}|^2 \Big].
\end{aligned}
\end{equation}
Accordingly, the normalized second-order intensity correlation function of the mixed thermal field is
\begin{equation}\label{eq:mixg2}
\begin{aligned}
g^{(2)}_m &= \frac{\left\langle  I_{1m}I_{2m} \right\rangle}{\bar I_{1m} \bar I_{2m} }\\
&=1\!+\!\frac{\sum_{i}\alpha_i^4\!+\!\sum_{k}\beta_k^4\!+\!2\cos^2\theta\sum_{i,k}\alpha_i^2\beta_k^2|K_{ik}|^2}{\big(\sum_{i} \alpha_i^2 + \sum_{k} \beta_k^2\big)^2}\\
&=1\!+\!\frac{1}{M_m},
\end{aligned}
\end{equation}
where
\begin{equation}\label{eq:mixnum}
\begin{aligned}
M_m &= \frac{\big( \bar I_1 +\bar I_2 \big)^2}{\sum_{i}\alpha_i^4\!+\!\sum_{k}\beta_k^4\!+\!2\cos^2\theta\sum_{i,k}\alpha_i^2\beta_k^2|K_{ik}|^2}
\end{aligned}
\end{equation}
denotes the average mode number of the mixed thermal field. From Eq.~(\ref{eq:mixnum}), one sees that although the intensity of each field $\bar I_1=\sum_{i} \alpha_i^2$, $\bar I_2=\sum_{k}\beta_k^2$ are factors influencing $M_m$, but the key term determining $M_m$ is the interference term $\cos^2\theta\sum_{i,k}\alpha_i^2\beta_k^2|K_{ik}|^2$, in which the coefficients $\theta$ and $K_{ik}$, used to describe the overlap of polarization and mode structures of TMs between $E_1$ and $E_2$, play an important role. To better understand the mode property of the mixed thermal field, we analyze the dependence of mixed field $g_m^{(2)}$ in Eq.~(\ref{eq:mixg2}) in the following three cases.

In the first case, the mode structures and polarization of the two independent fields $E_1$ and $E_2$ are identical, the interference term $\cos^2\theta\sum_{i,k}\alpha_i^2\beta_k^2|K_{ik}|^2$ in Eqs.~(\ref{eq:mixg2}) and (\ref{eq:mixnum}) takes the maximum value, i.e., we have $K_{i,k}=\delta_{i,k}$  and $\cos\theta = 1$. Under such conditions, Eq.~(\ref{eq:mixg2}) has the simplified form
\begin{equation}\label{eq:firstcase}
\begin{aligned}
g^{(2)}_m &= 1+\frac{\sum_{i}(\alpha_i^2\!+\!\beta_i^2)^2 }{\big[ \sum_i(\alpha_i^2+\beta_i^2)\big] ^2}.
\end{aligned}
\end{equation}
For $\beta_i=0$ or $\alpha_i=0$, $g^{(2)}_m$ becomes the same as $g^{(2)}$ for individual field $E_1$ or $E_2$, i.e., $g^{(2)}_m=g^{(2)}_1=1+\frac{1}{M_1}$ or
$g^{(2)}_m=g^{(2)}_2=1+\frac{1}{M_2}$, 
which are in consistent with the photon statistics for a thermal field (see Eq.~(\ref{eq:singleg2})). $M_1$ and $M_2$ are the mode numbers of field $E_1$ and $E_2$, respectively. Since it is difficult to obtain a analytical solution form the general expression in Eq.~(\ref{eq:firstcase}), we assume $M_1 \leq M_2$ and the intensities for both $E_1$ and $E_2$ are equally distributed in each TM, i.e., 
\begin{equation}\label{eq:approx}
\begin{aligned}
\alpha_i= \alpha \;(i = 1,\cdots,M_1),\\
\beta_k= \beta \;(k = 1,\cdots,M_2). 
\end{aligned}
\end{equation} 
Under the assumptions in Eq.~(\ref{eq:approx}), Eq.~(\ref{eq:firstcase}) can be approximated as
\begin{equation}\label{eq:upperbound}
\begin{aligned}
g^{(2)}_m &=1\!+\!\frac{M_1(\alpha^2+\beta^2)^2+(M_2-M_1)\beta^4
}{\left( M_1\alpha^2+ M_2\beta^2\right) ^2}\\
&= 1\!+\!\frac{\mathcal{R}^2}{M_1}\!+\!\frac{1-\mathcal{R}^2}{M_2},
\end{aligned}
\end{equation}
with
\begin{equation}\label{eq:ratio}
\mathcal{R} = \frac{\bar I_1}{\bar I_1+\bar I_2},
\end{equation}
where ${\bar I}_1=M_1 \alpha^2$ and ${\bar I}_2=M_2 \beta^2$, and $\mathcal{R}$ is the ratio between the intensity of field $E_1$ and the total intensity of two fields. Eq.~(\ref{eq:upperbound}) can be viewed as the upper bound for the interference shown up in the mixed field. When $M_1=M_2$, the mode of the mixed filed is the same as $E_1$ and $E_2$ and is irrelevant to $\mathcal{R}$, i.e., $g^{(2)}_m=g^{(2)}_1=g^{(2)}_2$ always holds. When $M_1< M_2$, however, $g^{(2)}_m$ decreases from $g_1^{(2)}$ to $g_2^{(2)}$ with the decrease of $\mathcal R$ and reaches the minimum $g^{(2)}_m=g_2^{(2)}$ at $\mathcal R=0$.

In the second case, the modes of the two independent fields $E_1$ and $E_2$ are orthogonal, the interference term  $\cos^2\theta\sum_{i,k}\alpha_i^2\beta_k^2|K_{ik}|^2$ in Eq.~(\ref{eq:mixg2}) takes the minimum value $0$, which means the polarization of the two filed are perpendicular to each other or there is no overlap between the TMs of $E_1$ and $E_2$, i.e., we have $K_{i,k}=0$ or $\cos\theta =0$. Under such conditions, Eq.~(\ref{eq:mixg2}) is simplified as
\begin{equation}\label{eq:secondcase}
g^{(2)}_m = 1 + \frac{\sum_{i}\alpha_i^4+\sum_{k}\beta_k^4}{(\sum_{i}\alpha_i^2+\sum_{k}\beta_k^2)^2}.
\end{equation}
By assuming $M_1\leq M_2$ and taking the assumptions in Eq.~(\ref{eq:approx}), Eq.~(\ref{eq:secondcase}) can be approximated as
\begin{equation}\label{eq:lowerbound}
g^{(2)}_m = 1 + \frac{M_1\alpha^4+M_2\beta^4}{(M_1\alpha^2+M_2\beta^2)^2}\\
=1 + \frac{\mathcal{R}^2}{M_1}+\frac{(1-\mathcal{R})^2}{M_2},
\end{equation}
which can be viewed as the lower limit for the interference shown in mixed field. It is straightforward to deduce intensity correlation function of mixed field drops to the minimum $g^{(2)}_{min}=1+\frac{1}{M_1+M_2}$ 
when the relative ratio of the two fields takes the value of $\mathcal{R}=\frac{M_1}{M_1+M_2}$. In particular, if $M_1=M_2=M$, the minimum $g^{(2)}_m$ of the mixed field, $g^{(2)}_{min}=1+\frac{1}{2M}$, is obtained for the two fields with equal intensity, i.e., $\mathcal{R}=0.5$.

In the third case, we have $0 < K_{i,k} < 1$ and $\cos\theta =1$. This is the most general situation for two independent thermal fields, because it is vary easy to realize the perfect matching for polarization, however, it is impossible to perfectly match the TMs of two fields with $M_{1,2}>1$ unless their emitting light sources are identical. Under this condition, by defining
\begin{eqnarray}\label{eq:V}
{\cal V} \equiv\frac{\sum_{i,k}\alpha_i^2\beta_k^2|K_{ik}|^2}{\sum_{i}\alpha_i^2\beta_i^2},
\end{eqnarray}
 Eq.~(\ref{eq:mixg2}) can be rewritten as
\begin{eqnarray}\label{eq:thirdcase}
g_m^{(2)} = 1 + \frac{\sum_{i}\alpha_i^4\!+\!\sum_{k}\beta_k^4\!+\!2{\cal V}\sum_{i}\alpha_i^2\beta_i^2}{\big(\sum_{i} \alpha_i^2 + \sum_{k} \beta_k^2\big)^2}.
\end{eqnarray}
Moreover, by assuming $M_1 \leq M_2$ and taking the assumptions in Eq.~(\ref{eq:approx}), Eq.~(\ref{eq:thirdcase}) can be approximated as
\begin{equation}\label{eq:midg2}
\begin{aligned}
g_m^{(2)} &=1+ \frac{M_1\alpha^4\!+\!M_2\beta^4\!+\!2M_1{\cal V}\alpha^2\beta^2}{\big(M_1\alpha^2 + M_2 \beta^2\big)^2}\\
&=1+ \frac{\mathcal{R}^2}{M_1}\!+\!\frac{(1-\mathcal{R})(1-\mathcal{R}+2{\cal V}\mathcal{R})}{M_2},
\end{aligned}
\end{equation}
with
\begin{eqnarray}\label{eq:norV}
{\cal V} = {1\over M_1}\sum_{i=1,k=1}^{M_1,M_2}|K_{ik}|^2.
\end{eqnarray}
Because $\phi_i(t)$ and $\varphi_k(t)$ are complete bases of TMs for the thermal fields of $E_1$ and $E_2$, respectively, $K_{ik}$ in Eq.~(\ref{eq:norV}) is their transition matrix element which must satisfy $\sum_{k=all} |K_{ik}|^2 = 1$ so that
\begin{eqnarray}\label{eq:Vrelation}
\sum_{k=1}^{M_2}|K_{ik}|^2\le \sum_{k=all} |K_{ik}|^2 = 1.
\end{eqnarray}
This leads to ${\cal V}\le 1$. For the extreme cases, as we have discussed in the first case and second case, we have $K_{ik} = \delta_{ik}$ and $K_{ik} = 0$ ($i=1,\cdots,M_1;k=1,\cdots,M_2$), respectively, Eq.~(\ref{eq:midg2}) has the simplified forms, which are exactly the upper bound and lower limit in Eqs.~(\ref{eq:upperbound}) and (\ref{eq:lowerbound}), respectively. For the general case, the TMs of $E_1$ and $E_2$ are partially overlapped, we have $0<{\cal V} <1$. From  Eqs.~(\ref{eq:norV}) and (\ref{eq:Vrelation}), one sees that the value of $\cal V$ depends on $M_1$, $M_2$, and the details of the mode excitation through $K_{ik}$-quantities. Notice that Eqs.~(\ref{eq:midg2})-(\ref{eq:Vrelation}) are approximations under the assumption of Eq.~(\ref{eq:approx}),
which is usually not the case for real thermal fields, so the experimental results presented in Sec.\;III.\;B may only qualitatively agree with Eq.~(\ref{eq:midg2}) when the general case of partial mode overlapping $0<{\cal V} <1$ is verified.

\section{Experiment}

\begin{figure*}[!htp]
	\includegraphics[width=15cm]{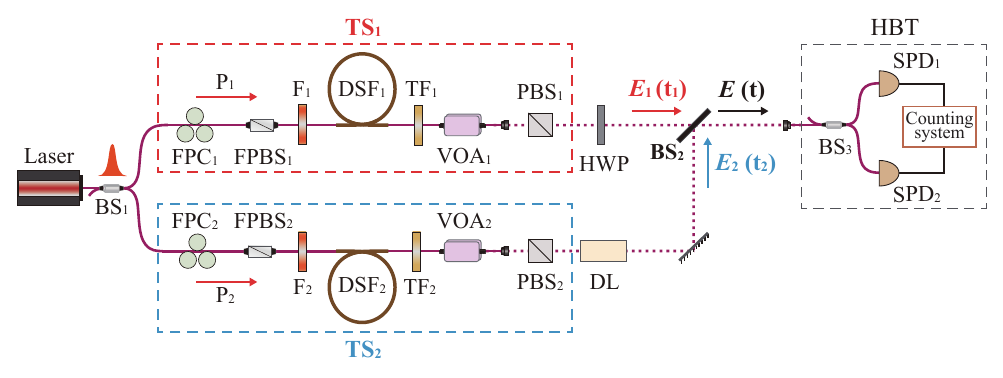}
	\caption{\label{setup} Experimental setup for verifying the interference between two multi-mode thermal fields $E_1$ and $E_2$. P$_1$-P$_2$, Pump; DSF$_1$-DSF$_2$, dispersion shifted fiber; F$_1$-F$_2$, filter; TF$_1$-TF$_2$, tunable filter; FPC$_1$-FPC$_2$, fiber polarization controller; FPBS$_1$-FPBS$_2$, fiber polarization beam splitter; PBS$_1$-PBS$_2$; polarization beam splitter; VOA$_1$-VOA$_2$, variable optical attenuator; HWP, half wave plate; BS$_1$-BS$_3$, 50/50 beam splitter; DL, delay line; SPD$_1$-SPD$_2$, single photon detector. The solid lines and dotted lines in the scheme respectively denote the optical fiber propagation and free space propagation.}
\end{figure*}

We perform a few experiments to verify the theoretical results obtained in the previous section. The experimental setup is shown in Fig.~\ref{setup}. The mixed field $E(t)$ is obtained by coupling two multi-mode thermal fields $E_1(t_1)$ and $E_2(t_2)$ with a 50/50 beam splitter (BS$_2$). The thermal fields $E_1$ and $E_2$ are respectively originated from two independent thermal sources (TSs). Each TS is based on the radiation of nonlinear process excited in dispersion shifted fiber (DSF) with a pulsed pump. The nonlinear process in DSF is either spontaneous Raman scattering (SRS) or spontaneous four wave mixing (SFWM). Both the stocks wave of SRS and the individual signal field of SFWM are in thermal state ~\cite{Li05c,Ma-pra11}. SRS with a broad gain bandwidth always occurs whenever the strong pump is propagating along optical fibers~\cite{book-Agrawal}. However, the SFWM, which dominates the SRS, occurs only when the phase matching condition in DSF is satisfied~\cite{book-Agrawal}. Moreover, the basis of TMs for each TS is determined by the specific nonlinear process in the DSF and its pulsed pump field~\cite{li08a-opex,Liu-OE16}. $E_1$ and $E_2$ with same temporal mode structure can only obtained when the excitation conditions of TS$_1$ and TS$_2$ are exactly the same. When the details of nonlinear process in DSF is changed, which include the dispersion of DSF or the pump wavelength and bandwidth, the TMs basis of TS will be accordingly changed.

The two polarized pumps, P$_1$ and P$_2$, respectively used for pumping DSF$_1$ and DSF$_2$, are achieved by dividing the output of a mode locked fiber laser into two with a 50/50 beam splitter (BS$_1$). The repetition rate of the laser is about 36.88 MHz. The central wavelengths of the two Gaussian shaped pumps $P_1$ and $P_2$ are the same, but their bandwidth are respectively determined by that of the filter F$_{1}$ and F$_{2}$. The pulse duration of $P_{1,2}$ can be adjusted to be $\sim$ 10 ps or $\sim$ 5 ps, which is achieved by setting the bandwidth of $F_{1,2}$ to 0.4 or 1 nm. The power of P$_1$ (P$_2$) is adjustable by using the combination of fiber polarization controller FPC$_1$ (FPC$_2$) and fiber polarization beam splitter FPBS$_1$ (FPBS$_2$).

The output of TS$_{1,2}$, originated from the nonlinear process in DSF$_{1,2}$, propagates through a tunable filter TF$_{1,2}$ to separate the thermal field $E_{1,2}$ from the residual pump of P$_{1,2}$ and to reshape the bandwidth of $E_{1,2}$. It is well known that when $E_{1,2}$ can be viewed as in single mode and the coherence time of $E_{1,2}$ is much longer than the pulse duration of TS$_{1,2}$, the spectrum of TF$_{1,2}$ defines the mode property of $E_{1,2}$ \cite{Ou97}. However, in this paper, we are mainly interested in $E_{1}$ and $E_{2}$ with mode number $M_{1,2} > 1$. So the tunable filter TF$_{1,2}$ will has no effect on TMs basis for $E_{1,2}$ with $M_{1,2} > 1$, but changes the coefficients $\alpha_i$ and $\beta_k$ of the specified TMs $\phi_i(t)$ and $\varphi_k(t)$. The pulse duration of each thermal field $E_{1,2}$, determined by the pump pulse width and bandwidth of TF$_{1,2}$, is within tens picoseconds. The central wavelength of $E_{1,2}$ is the same as that of TF$_{1,2}$ and can be tuned within the telecom band. The mode number of $E_{1,2}$ can be conveniently changed by varying the FWHM of TF$_{1,2}$, which can be adjusted from 0.3 to 2.5 nm. The polarization state of $E_{1,2}$ is defined by the polarization beam splitter PBS$_{1,2}$, and the relativel polarization angle of $E_{1}$ and $E_{2}$ can be changed by a half wave plate (HWP). The relative strength of $E_{1}$ and $E_{2}$, described by the ratio $\cal R$ in Eq.~(\ref{eq:ratio}), is changed by using the variable optical attnuators VOA$_1$ and VOA$_2$.

By launching fields $E_1$ and $E_2$ respectively into the two input ports of a 50/50 beam splitter (BS$_2$), we obtain the combined field $E(t)=\frac{1}{ \sqrt{2}}(E_1(t)+E_2(t+\tau))$ at one output of BS$_2$, where $\tau=t_2-t_1$. Here the relative delay $\tau$ between $E_1$ and $E_2$ is introduced by passing the field $E_2$ through a delay line (DL). The interference effect in the combined field $E(t)$ are characterized by using a Hanbury Brown-Twiss (HBT) interferometer, consisting of a 50/50 BS (BS$_3$) and two single photon detectors, SPD$_1$ and SPD$_2$. The two SPDs (InGaAs-based) are operated in a gated Geiger mode. The 2.5-ns gate pulses on SPDs arrive at a repetition rate the same as that of the laser, and the dead time of the gate is set to be 10 $\mu$s. The intensity correlation function is measured when the optical path lengths from BS$_2$ to SPD$_1$ and SPD$_2$ are equal. During the measurement, the counting system records the individual count rate of SPD$_1$ and SPD$_2$, $N_1$ and $N_2$, which are proportional to the intensity of detected fields. In the meantime, the two-fold coincidence rate of SPD$_1$ and SPD$_2$, $N_c$, which reflects the correlation of two detected fields, is recorded as well. $g_m^{(2)}$ of the mixed field is then obtained from the relation $g_m^{(2)}={N_c}/{\big(N_1N_2\big)}$. In the experiment, the mode number of the thermal field $E_1$ or $E_2$ is characterized by directly sending the individual field into the HBT and measuring its intensity correlation $g_{1,2}^{(2)}$. The mode number $M_{1,2}$ is deduced through the relation $g_{1,2}^{(2)}=1+1/M_{1,2}$.

\subsection{Mixing of two thermal sources of identical mode structure}

We first verify the upper bound and lower limit for the interference effect shown in the mixed field by using the TSs with same mode profiles, which means the mode bases of $E_1$ and $E_2$, $\phi_i(t)$ and $\varphi_k(t)$, are perfectly overlapped, i.e, $\phi_i(t)=\varphi_i(t)$. In these experiments, the thermal sources for emitting $E_1$ and $E_2$ are the same in every detail. To ensure the exact similarity, the two TSs are based on the identical Raman scattering process respectively occurred in two DSFs. In experiment, the two DSFs, with length and zero dispersion wavelength of 300 m and 1552 nm, are identical, and the central wavelength of two pumps is selected to be 1541 nm, at which the phase matching condition of SFWM in the DSF is not satisfied. Moreover, the FWHM and average power for both P$_1$ and P$_2$ are 1 nm and 1 mW, respectively.
\begin{figure}[!htp]
	\includegraphics[width=8.5cm]{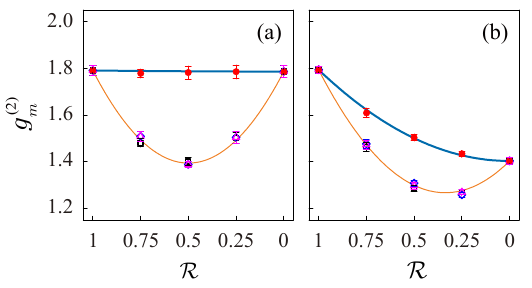}
	\caption{\label{fig:4} Intensity correlation function $g_m^{(2)}$ of the mixed thermal field when the mode numbers of the two independent fields $E_1$ and $E_2$ are (a) $M_1=M_2=1.25$ and (b) $M_1=1.25, M_2=2.5$, respectively. The solid circles are obtained when the mode profiles of $E_1$ and $E_2$ satisfy the upper bound conditions, $K_{ik}=\delta_{ik}$ and $\cos \theta=1$, while others (squares, diamonds and triangles) are obtained when the mode profiles of $E_1$ and $E_2$ satisfy the lower limit conditions , $K_{ik}=0$ or $\cos \theta=0$. The thick and thin curves are the results calculated by substituting mode numbers of $E_1$ and $E_2$ into upper bound and lower limit of $g^{(2)}_m$ in Eqs.~(\ref{eq:upperbound}) and (\ref{eq:lowerbound}), respectively. }
\end{figure}

In the experiment of verifying the upper bound of two-photon interference in mixed field, the central wavelengths of both TF$_1$ and TF$_2$ are 1564 nm, and relative delay is set to $\tau=0$ by carefully adjusting the DL. Moreover, the polarization for $E_1$ and $E_2$ at BS$_2$ are adjusted to be the same. Hence, the conditions $K_{i,k}=\delta_{i,k}$ and $\cos\theta=1$ (in Eq.~(\ref{eq:mixg2})) are satisfied. We conduct the measurement of $g_m^{(2)}$ for the mixed field $E(t)$ when the bandwidth of TF$_1$ is fixed at 0.75 nm but bandwidth of TF$_2$ is 0.75 and 2.25 nm, respectively. In the two cases, the mode numbers of $E_1$ and $E_2$ are (i) $M_1=M_2=1.25$, and (ii) $M_1=1.25$, $M_2=2.5$. The data of $g_m^{(2)}$ for cases (i) and (ii) with ${\cal R} =0.75,0.5, 0.25$ is shown as the solid circles in Figs.~\ref{fig:4}(a) and \ref{fig:4}(b) respectively. We calculate the result of $g_m^{(2)}$ as a function of ${\cal R}$ by substituting the experimental parameters into Eq.~(\ref{eq:upperbound}), as shown as the thick solid curves in Figs.~\ref{fig:4}(a) and \ref{fig:4}(b), respectively. The results indicate the theory prediction for the upper bound fit the experimental data very well.

In the experiment of verifying the lower limit of two-photon interference shown in mixed field, we set the experimental parameters the same as those for verifying the upper bound, except the polarization states of $E_1$ and $E_2$ in front of the BS$_2$ are orthogonal ($\vec{e}_1\perp \vec{e}_2$) or the delay between $E_1$ and $E_2$ at the BS$_2$ is adjusted by DL so that the approximation $\tau\rightarrow \infty$ is valid. Therefore, the condition of $K_{i,k}=0$ or $\cos\theta=0$ in Eq.~(\ref{eq:mixg2}) is satisfied. We then conduct the measurement of $g_m^{(2)}$ of the mixed field under the condition of $\vec{e}_1\perp \vec{e}_2$ or $\tau\rightarrow \infty$ when the relative strength of the two thermal fields is ${\cal R}=0.75,0.5, 0.25$. In the measurement, the results for the two kinds of mode number combinations for $E_1$ and $E_2$, the same as cases (i) and (ii) for verifying the upper bound, are shown in Fig.~\ref{fig:4}. The data obtained under of the condition of $\vec{e}_1\perp \vec{e}_2$ and $\tau\rightarrow \infty$, respectively, is represented by the hollow squares and diamonds in Figs.~\ref{fig:4}(a) and 4(b). As a comparison, we substitute the experimental parameters into Eq.~(\ref{eq:lowerbound}) to calculate the corresponding $g_m^{(2)}$ by varying ${\cal R}$, as shown as the thin solid curves in Figs.~\ref{fig:4}(a) and \ref{fig:4}(b), respectively. It can be seen that the theory prediction of the lower limit perfectly agrees with the experimental results. Additionally, we also achieve $K_{ik}=0$ by adjusting the central wavelength of $E_1$ field from 1564 nm to 1566 nm using TF$_1$, which leads to the orthogonality of $E_1$ and $E_2$ as well. By keeping the other parameters the same as those for testing the upper bound, we measure $g_m^{(2)}$ for ${\cal R}=0.75,0.5, 0.25$. The data for the mode number the same as cases (i) and (ii) is shown as the hollow triangles in Figs.~\ref{fig:4}(a) and \ref{fig:4}(b), respectively, which overlap with the hollow squares and diamonds and well fit the theory curves for lower limit.

\begin{figure}[!htp]
	\includegraphics[width=8.5cm]{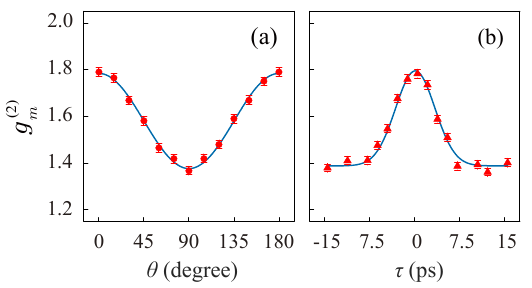} 
	\caption{\label{fig:5} Intensity correlation function $g_m^{(2)}$ measured by (a) varying the angle of the polarization $\theta$ between linearly polarized $E_1$ and $E_2$  and by (b) varying the delay $\tau$ between $E_1$ and $E_2$. In the measurement, $M\!=M_1\!=\!M_2\!=\!1.25$ $ (g^{(2)}_m\!=\!g_1^{(2)}\!=\!g_2^{(2)}\!=\!1.8)$ and ${\cal R}\!=\! 0.5$. The solid curves in (a) and (b) are obtained by substituting experimental parameters in the formula $g_m^{(2)} =1+\frac{1}{2M}(1+\cos^2\theta)$ and $g_m^{(2)} =1+\frac{1}{2M}\{1+\exp[-\tau^2\sigma^2(g^{(2 )}-1)^2/2]\}$, respectively. }
\end{figure}

We next study the influence of mode match effect on interference by changing $K_{ik}$ or $\cos \theta$ in Eq.~(\ref{eq:mixg2}), which can be achieved by adjusting the delay or polarization angle between the linearly polarized fields $E_1$ and $E_2$, while other experimental parameters are the same as those for verifying upper bound. In this experiment, the intensities for both $E_1$ and $E_2$ are the same, and the mode numbers for both $E_1$ and $E_2$ are fixed at $1.25$. The solid circles in Fig.~\ref{fig:5}(a) present measured $g_m^{(2)}$ as a function of the angle $\theta$  between the polarization of $E_1$ and $E_2$ under the condition of $\tau=0$. Moreover, we calculate $g_m^{(2)}$ by substituting the experimental parameters, ${\cal R}\!=\! 0.5$ and $M\!=M_1\!=\!M_2\!=\!1.25$, into the formula $g_m^{(2)}=1+\frac{1}{2M}(1+\cos^2\theta)$ (a simplified form of Eq.~(\ref{eq:mixg2})), which agrees well with the experimental data. The triangles in Fig.~\ref{fig:5}(b) present the measured $g_m^{(2)}$ as a function of relative delay $\tau$ under the condition of $\vec{e}_1\cdot \vec{e}_2=1$. Since the mode profiles of $E_1$ and $E_2$ are the same, the effect of delay can be calculated by using Eq.~(13) in Ref.~\cite{Ma-pra11}. In this way, we have $g_m^{(2)}=1+\frac{1}{2M}\{1+\exp[\frac{-\tau^2\sigma^2(g^{(2)}-1)^2}{2}]\}$ with $\sigma$ denoting the spectral bandwidth of thermal fields $E_1$ and $E_2$. Substituting the experimental parameters into the formula, we obtain the solid curve in Fig.~\ref{fig:5}(b), which is well fitted with the experimental results.

\subsection{Mixing of two thermal sources of different mode structure}

The results in Figs.~\ref{fig:4} and \ref{fig:5} demonstrate that our theoretical analysis are correct for fields $E_1$ and $E_2$ emitted by sources that are independent but have the same temporal mode structures. In nature, thermal sources from different kind of sources usually have different mode structure. In this case, $0<K_{ik}<1$, even for $E_1$ and $E_2$ with zero delay ($\tau=0$), perfect polarization states $\vec{e}_1\cdot \vec{e}_2=1$, and identical spectrum. In this subsection, we will perform experiments when the TMs of $E_1$ and $E_2$ are partially overlapped. To obtain the two thermal sources with different mode profiles, we proceed with two approaches.

\begin{figure}[!htp]
	\includegraphics[width=8.5cm]{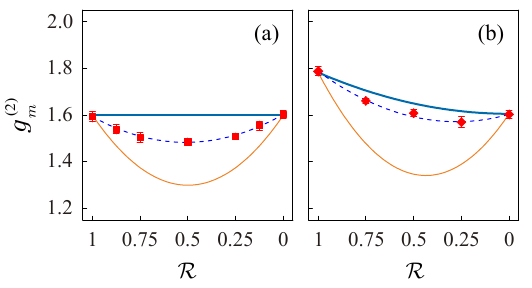} 
	\caption{\label{fig:6} Intensity correlation function $g_m^{(2)}$ for the two independent multi-mode thermal fields $E_1$ and $E_2$ generated from two RS processes with different mode profiles. In the experiment, the conditions $0<K_{i,k}<1$ and $\cos \theta = 1$, $\tau=0$ are satisfied. The squares and diamonds are obtained for $E_1$ and $E_2$ with mode numbers of (a) $M_1=M_2=1.67$ and (b) $M_1=1.25,\; M_2=1.67$, respectively. The dashed curves are fittings of Eq.~(\ref{eq:midg2}) with ${\cal V} = 0.62$ for (a) and 0.82 for (b). The thick and thin curves are the theory predictions of upper bound and lower limit, calculated by substituting mode numbers of $E_1$ and $E_2$ into Eqs.~(\ref{eq:upperbound}) and (\ref{eq:lowerbound}), respectively.
	}
\end{figure}

The first approach still utilizes Raman scattering as the two thermal sources but the pump bandwidths are different, resulting in different temporal profiles for the generated Raman scattering. In this case, the experimental parameters are the same as those for verifying the upper bound, but the bandwidths of the two pump fields, P$_1$ and P$_2$, are set to 0.4 nm and 1.0 nm, respectively. Moreover, $E_1$ is obtained by passing through TF$_1$ with FWHM of 0.5 nm, which corresponds to the average mode number $M_1=1.67$ ($g_1^{(2)}=1.6$); while $E_2$ is obtained by passing through TF$_2$ with FWHM of 0.75 nm or 1.3 nm, which correspond to the average mode number of $M_2=1.25$ or $M_2=1.67$. For $E_1$ and $E_2$ with the two kind of mode number combinations, we then measure $g_m^{(2)}$ of the mixed field $E(t)$ when the relative strength of $E_1$ and $E_2$ is ${\cal R}=0.75, 0.5, 0.25$, as shown by the solid circles in Figs.~\ref{fig:6}(a) and \ref{fig:6}(b), respectively. We fit the data by using Eq.~(\ref{eq:midg2}) (dashed curves) with a best fitting value of ${\cal V} =0.62$ for Fig.~\ref{fig:5}(a) and ${\cal V} =0.82$ for Fig.~\ref{fig:5}(b). As a comparison, we also plot upper bound (thick curves) and lower limit (thin curves) for $E_1$ and $E_2$ with the given combination of mode numbers as a function of $\cal R$, by substituting the parameters into Eqs.~(\ref{eq:upperbound}) and (\ref{eq:lowerbound}), respectively. It is clear that the experimental data of $g_m^{(2)}$ is within the upper bound and lower limit of interference in mixed field due to the partial mode overlap between the TMs basis of $E_1$ and $E_2$. The fitting parameter $\cal V$ is within the range of $0 < {\cal V} < 1$, which qualitatively agree with the theory prediction in Eq.~(\ref{eq:midg2}).

\begin{figure}[!htp]
	\includegraphics[width=10.5cm]{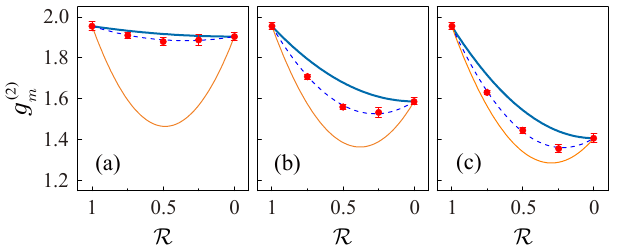} 
	\caption{\label{fig:7} Intensity correlation function $g_m^{(2)}$ for the two independent multi-mode thermal fields $E_1$ and $E_2$ when the mode number $M_1$ is fixed at 1.05, but $M_2$ is (a) 1.11, (b) 1.67, (c) 2.5, respectively. The data is fitted to Eq.~(\ref{eq:midg2}) (dashed curves) with the best fit value ${\cal V}$ = 0.93 for (a), 0.62 for (b), 0.48 for (c), respectively. In the measurement, $E_1$ and  $E_2$ are originated from the SFWM and SRS in DSFs, respectively, and the condition $0 < K_{ik} <1$ and $\theta=0, \tau=0$ are satisfied. The thick and thin curves are the results calculated by substituting mode numbers of $E_1$ and $E_2$ into Eqs.~(\ref{eq:upperbound}) and (\ref{eq:lowerbound}), respectively.
	}
\end{figure}

The second approach is to switch the thermal source TS$_1$ to SFMW in DSF and keep the other, TS$_2$, the same as before. In this case, TS$_1$ is replaced by using another DSF having zero dispersion wavelength at 1540 nm. With this replacement, the phase matching condition of SFWM in DSF$_1$ is satisfied. Moreover, we increase the pump power P$_1$ from 1 mW to 2 mW, so that the gain of four wave mixing is quite high and the intensity of Raman scattering (RS) in DSF$_1$ is negligible \cite{Liu-OE16}. As a result, up to 98$\%$ photons in thermal field $E_1$ are originated from the individual signal field of SFWM while the field $E_2$ from DSF$_2$ is still radiated by Raman scattering. So mode profiles of TS$_1$ and TS$_2$ are totally different. Moreover, $E_1$ is obtained by passing through TF$_1$ with FWHM of 0.3 nm, which corresponds to average mode number $M_1=1.05$ ($g_1^{(2)}=1.95$), while $E_2$ is obtained by passing through TF$_2$ with FWHM of 0.5, 1.5 and 2.5 nm, which correspond to the mode number of $M_2=1.11$, $1.67$ and $2.5$, respectively. For $E_1$ and $E_2$ with the three kinds of mode number combinations, we then measure $g_m^{(2)}$ of the mixed field $E(t)$ when the relative strength is ${\cal R}=0.75, 0.5, 0.25$, as shown by the solid circles in Figs.~\ref{fig:7}(a), \ref{fig:7}(b) and \ref{fig:7}(c), respectively. The data is fitted to Eq.~(\ref{eq:midg2}) (dashed curves) with the best fit value ${\cal V}$ of 0.93, 0.62 and 0.48, respectively. Additionally, according to the three kinds of mode number combinations of $M_1$ and $M_2$, we calculate the upper and lower bounds of $g_m^{(2)}$ as a function of ${\cal R}$ by using Eqs.~(\ref{eq:upperbound}) and (\ref{eq:lowerbound}), as shown by the thick and thin solid curves in Figs. \ref{fig:7}(a), \ref{fig:7}(b) and \ref{fig:7}(c), respectively. Simiar to Fig.~\ref{fig:6}, the experimental results of $g_m^{(2)}$ in Fig.~\ref{fig:7} are within the upper and lower bounds, which agree with the theory prediction in Eq.~(\ref{eq:midg2}). Moreover, we notice that in this experiment, $E_1$ field is very close to single mode, so its mode profile here can be approximated by the spectrum of TF$_1$ \cite{Ou97,Ma-pra11}. In this case, it is reasonable that the fitting parameter ${\cal V}$, reflecting the degree of mode mismatching between $E_1$ and $E_2$, decreases with the increase of $M_2$.
\vspace{-1.5em}
\section{Conclusions and discussion}

We have developed a general theory for analyzing the mode profile of a field formed by mixing two multi-mode thermal fields, and found that the two-photon interference between the two independent fields play an important role. Comparing with $g^{(2)}$ for one of the individual field with less average mode number, $g_m^{(2)}$ of the mixed field always decreases, but the amount of drop depends on the relative overlap between the mode structures of the two thermal fields and their relative strength. Although the analytical expression of $g_m^{(2)}$ is deduced under a rough assumption of Eq.~(\ref{eq:approx}), we find that the measured $g_m^{(2)}$ well agree with the theory predictions no matter the modes of two multi-mode thermal fields involved in the interference are identical, orthogonal or partially overlapped, as long as the mode structures of two thermal fields are the same. On the other hand, when the mode structures of two thermal sources are not identical and the multi-mode thermal fields $E_1$ and $E_2$ are only partially overlapped, our experimental results qualitatively agree with the prediction in Eq.~(\ref{eq:midg2}). We believe this is because the assumptions in Eq.~(\ref{eq:approx}) used to deduce Eq.~(\ref{eq:midg2}) from Eq.~(\ref{eq:mixg2}) deviate from the thermal sources used in experiment. In order to precisely predict the theory curve of $g_m^{(2)}$ in this case, instead of using the general theory in Sec. II, we need to resort the specific model for describing the nonlinear process~\cite{Liu-OE16,li08a-opex,Garay-Palmett07,Walmsley09}, from which the details of TMs for each TS can be obtained and the accurate simulation of $g_m^{(2)}$ can be done~\cite{Ma-JOSAB-2015}. We believe our investigation is useful for analyzing the signals carried by the intensity correlation of thermal fields, such as improving the SNR of ghost imaging and analyzing the mode property of multi-mode quantum field \cite{ghost,brecht15}.

\begin{acknowledgments}
This work was supported in part by National Natural Science Foundation of China (11527808, 91736105), the National Key Research and Development Program of China (2016YFA0301403), 973 program of China (2014CB340103), and by the 111 project B07014.
\end{acknowledgments}


\nocite{*}


\end{document}